\documentclass[12pt]{article}

\usepackage{amsmath,amssymb,graphicx} 
\usepackage{epsf}
\usepackage{pstricks}
\usepackage{cite}

\newcommand{\beq}{\begin{eqnarray}}
\newcommand{\eeq}{\end{eqnarray}}

\newcommand{\centeron}[2]{{\setbox0=\hbox{#1}\setbox1=\hbox{#2}\ifdim
                                        
\wd1>\wd0\kern.5\wd1\kern-.5\wd0\fi
\copy0

\kern-.5\wd0\kern-.5\wd1\copy1\ifdim\wd0>\wd1
                                       \kern.5\wd0\kern-.5\wd1\fi}}
\newcommand{\ltap}{\>\centeron{\raise.35ex\hbox{$<$}}
                               {\lower.65ex\hbox{$\sim$}}\>}
\newcommand{\gtap}{\>\centeron{\raise.35ex\hbox{$>$}}
                               {\lower.65ex\hbox{$\sim$}}\>}

\newcommand\ZZ{\hbox{\zfont Z\kern-.4emZ}}
\font\zfont = cmss10 
\newcommand{\sfrac}[2]{{\textstyle\frac{#1}{#2}}}

\def\tv#1{\vrule height #1pt depth 5pt width 0pt}

\begin{document}
\begin{titlepage}
\begin{flushright}
{\tt hep-ph/0409126}
\end{flushright}

\vskip.5cm
\begin{center}
{\huge \bf  Curing the Ills of Higgsless Models: \\
the $S$ Parameter and Unitarity}
\vskip.1cm
\end{center}
\vskip0.2cm

\begin{center}
{\bf
{Giacomo Cacciapaglia}$^{a}$, {Csaba Cs\'aki}$^{a}$,
{Christophe Grojean}$^{b,c}$,\\
{\rm and}
John Terning\footnote{Address after Jan. 1, 2005,
Department of Physics, University of California, Davis, CA  
95616.}$^{d}$}
\end{center}
\vskip 8pt

\begin{center}
$^{a}$ {\it Institute for High Energy Phenomenology\\
Newman Laboratory of Elementary Particle Physics\\
Cornell University, Ithaca, NY 14853, USA } \\
\vspace*{0.1cm}
$^{b}$ {\it Service de Physique Th\'eorique,
CEA Saclay, F91191 Gif--sur--Yvette, France} \\
\vspace*{0.1cm}
$^{c}$ {\it Michigan Center for Theoretical Physics,
Ann Arbor, MI 48109, USA}\\
\vspace*{0.1cm}
$^{d}$ {\it Theory Division T-8, Los Alamos National Laboratory, Los
Alamos,
NM 87545, USA} \\
\vspace*{0.3cm}
{\tt  cacciapa@mail.lns.cornell.edu, csaki@lepp.cornell.edu,
grojean@spht.saclay.cea.fr, terning@lanl.gov}
\end{center}

\vglue 0.3truecm

\begin{abstract}
\vskip 3pt
\noindent
We consider various constraints on Higgsless models of electroweak
symmetry breaking
based on a bulk SU(2)$_L\times$SU(2)$_R\times$U(1)$_{B-L}$ gauge group
in warped space.
First we show that the $S$ parameter which is positive if fermions are
localized
on the Planck brane can be lowered (or made vanishing) by changing the
localization of
the light fermions. If the wave function of the light fermions is
almost flat
their coupling to the gauge boson
KK modes will be close to vanishing, and therefore contributions to the
$S$ parameter
will be suppressed. At the same time the experimental bounds on such
$Z^\prime$ and $W^\prime$  gauge
bosons become very weak, and their masses can be lowered to make sure
that perturbative
unitarity is not violated in this theory before reaching energies of
several TeV.
The biggest difficulty of these models is to incorporate a heavy top
quark mass without
violating any of the experimental bounds on bottom quark gauge
couplings. In the simplest models of fermion masses a sufficiently
heavy top quark also implies an unacceptably large correction to the
$Zb\bar{b}$ vertex
and a large splitting between the KK modes of the top and bottom
quarks, yielding
large loop corrections to the $T$-parameter. We present possible
directions for model
building where perhaps these constraints could be obeyed as well.

\end{abstract}

\end{titlepage}

\newpage


\section{Introduction}
\label{sec:intro}
\setcounter{equation}{0}
\setcounter{footnote}{0}

The quest to uncover the origin of electroweak symmetry breaking has
been at the forefront
of particle physics for 25 years, and after a tremendous amount of
theoretical  effort it is clearer than
ever that we will need experiments to answer the question.  This
has been further emphasized by the explosion in the last few years
of a plethora of alternative electroweak symmetry breaking scenarios,
which bear little resemblance to the three traditional solutions: the
standard model (SM), the minimal supersymmetric standard model, and the
technicolor scenario. In addition to large extra
dimensions \cite{ADD}, warped extra dimensions \cite{RS}, gauge
component Higgses \cite{A5Higgs},  ``little" Higgses \cite{littleHiggs},
and ``fat"
Higgses \cite{fatHiggs}, one of the most
recent proposals, and in some ways most radical, is the Higgsless
scenario
\cite{CGMPT,CGPT,Nomura,BPR}.
These models take advantage of the fact that with a Higgs localized in
an extra dimension,
there exists
a limit where the Higgs decouples from
$WW$ scattering but with a finite $W$ mass.  The limit is achieved by
taking the Higgs vacuum
expectation value (VEV) to $\infty$. In this limit the gauge symmetry
breaking amounts to imposing
Dirichlet boundary conditions on the gauge field at one end of the
extra dimension \cite{CGMPT}.
Quarks and leptons can receive masses from boundary conditions as well
\cite{Nomura,BPR,CGHST}.
Since there is no contribution to $WW$ scattering from a Higgs boson,
these scattering amplitudes
are unitarized by another mechanism: exchanges of the Kaluza-Klein (KK)
tower of gauge
bosons \cite{CGMPT,otherunitarity}. Taking the extra dimension to be
anti-de Sitter (AdS) with an SU(2)$_L\times$SU(2)$_R\times$U(1)$_{B-L}$
gauge group in the bulk has the dual
advantage of raising the KK masses to phenomenologically acceptable
levels (and thus solving the ``little hierarchy problem'') and also
imposing a custodial isospin symmetry
which protects the ratio of the $W$ and $Z$ masses.
The presence of this custodial isospin symmetry follows from the
AdS/CFT correspondence \cite{ADMS}.
Further properties of Higgsless models and different variations
have been explored in
refs.~\cite{DHLR1,BN,CCGT,DHLR2,BPRS,MSU,corrections,Howard,Maxim,BRST,
spurion,Nick,OtherHiggsless,Papucci}.

This scenario shares many common features with technicolor models: in
fact
it can be thought of as a gravity dual of technicolor models, except
that there
are regions of parameter space where the theory is in fact weakly
coupled
and calculable. The leading corrections to electroweak precision
observables
have been calculated for the simplest setup in
refs.~\cite{DHLR1,BN,CCGT},
where
a large positive contribution to the $S$ parameter has been found. This
can
be lowered by introducing a brane induced kinetic term on the TeV brane
for the $B-L$ gauge group~\cite{CCGT},
however at the price of lowering the mass of
one of
the $Z^\prime$ modes to levels already excluded by LEP2~\cite{LEP}
and/or Tevatron\cite{Tevatron}.

In this paper we
point out that one can in fact easily eliminate the large contributions
to the $S$  parameter by changing the position of the light fermions.
The reason
behind this is simple: the oblique correction parameters $S,T,U$ on
their own are
meaningless until the normalization of the couplings between the
fermions and
the gauge bosons is fixed. An overall shift in the fermion gauge
boson couplings can be reabsorbed in the oblique correction parameters
and thus effectively change the predicted values of $S,T,U$. This is
exactly what happens when one changes the localization parameters of the
light fermions. Until now all calculations of the oblique parameters
in Higgsless models~\cite{DHLR1,BN,CCGT,DHLR2,BPRS} or
variations~\cite{MSU,corrections,Howard,Maxim} have
assumed that the fermions are strictly localized on the Planck brane,
in which
case one obtains a positive $S$ parameter. However, it has been known
for quite
a while~\cite{CET} that if fermions are localized on the TeV brane then
the $S$ parameter in the Randall-Sundrum model is in fact negative.
Therefore
it should be expected that there should be an intermediate position
where
$S$ exactly vanishes. This actually happens when the fermion wave
functions
are ``flat'', corresponding to localization parameter $c=1/2$. This is
just a simple consequence of the orthogonality of the KK mode wave
functions of the gauge bosons: when $c=1/2$  the coupling of the KK
gauge
bosons to the fermions vanishes, eliminating any possible additional
LEP or Tevatron constraints on this setup. The fact that for $c=1/2$
the $S$ parameter generically vanishes was first mentioned
in~\cite{ADMS}.

Another issue frequently discussed regarding Higgsless models is the
question
whether the first KK mode of the gauge bosons is light enough to
actually
unitarize the WW scattering amplitudes at weak
coupling~\cite{DHLR1,DHLR2,Papucci}. It follows  from general
arguments that the asymptotically growing terms in the individual
scattering amplitudes
always cancel, however one also needs to make sure that the finite
terms are
sufficiently small and in the perturbative regime, even after taking a
coupled channel
analysis into account. In this paper we point out
that by adjusting the value of the bulk curvature scale one can lower
the
masses of the KK gauge bosons to a few hundred GeV, without conflicting
the
direct search bounds due to the weak coupling of the light fermions to
the
KK modes, while the $S$ parameter can be made vanishing by adjusting the
localization parameter $c$ of the fermions. This way we show that the
two most commonly quoted problematic aspects of Higgsless models are in
fact
easily avoidable.

Rather, we find that the most serious issue of these models
is the inclusion of a heavy top quark. In the simplest implementation
of fermion masses one will generically either find a top quark mass
that is
too low, or a correction to the $Zb\bar{b}$ vertex that deviates
from the SM prediction beyond allowable levels. Furthermore, in most
cases when the top quark is sufficiently split from the bottom quark
there is
also a large splitting in the KK modes of the top and bottom quarks
leading to unacceptably large loop corrections to the $T$
parameter~\cite{ADMS}.
We present some speculations on possible extensions of the model
where the third generation could perhaps be included without violating
experimental bounds.

\section{The Model}
\label{sec:model}
\setcounter{equation}{0}

We will consider a bulk SU(2)$_L\times$SU(2)$_R\times$U(1)$_{B-L}$
gauge theory
on an AdS$_5$ background, working in the conformally flat metric
\beq
ds^2=  \left( \frac{R}{z}\right)^2   \Big( \eta_{\mu \nu} dx^\mu dx^\nu
- dz^2 \Big)
\eeq
where $z$ is on the interval $[R,R^\prime]$. The AdS curvature $R$ is
usually assumed to be of order $1/M_{Pl}$, however it is a freely
adjustable parameter.
In the following we will usually assume $R=10^{-19}$ GeV$^{-1}$ in all
the numerical examples, except in sec.~\ref{sec:unitarity}.
The parameter $R'$ sets the scale of the
gauge boson masses, and will therefore be $R' \sim 1/$TeV. As usual we
will
call the $z=R$ endpoint the Planck brane and $z=R'$ the TeV brane.
We will use the usual bulk Lagrangian, with canonically normalized
kinetic terms and in the unitary gauge, where all the $A_5$'s decouple
and we are left with a KK tower of vector fields,
$(A^{L}_{\mu},A^{R}_{\mu},B_{\mu})$~\cite{CGMPT,CGPT}.
We denote the 5D gauge couplings by $g_{5L}$, $g_{5R}$ and
$\tilde{g}_5$.
Electroweak symmetry breaking is achieved by the boundary conditions
that break $SU(2)_L \times SU(2)_R \rightarrow SU(2)_D$ on the TeV
brane and $SU(2)_R \times U(1)_{B-L} \rightarrow U(1)_Y$ on the Planck
brane.
We also consider kinetic terms allowed on the
branes~\cite{Nomura,CCGT}, that in terms of field stress tensors can be
parametrized:
\beq \label{genLagr}
\mathcal{L} = - \left[ \frac{r}{4} {W^L_{\mu \nu}}^2 +
\frac{r'}{4} {B^Y_{\mu \nu}}^2\right] \delta (z-R)
     - \frac{R'}{R} \left[ \frac{\tau'}{4} {B_{\mu\nu}}^2
+ \frac{\tau}{4} {W^D_{\mu \nu}}^2 \right]\delta (z-R')\,,
\eeq
where $A^D = (g_{5R} A^R + g_{5L} A^L)/\sqrt{g_{5R}^2+g_{5L}^2}$ and
$B^Y = (g_{5R} A^{R3} + \tilde g_5 B)/\sqrt{g_{5R}^2+ \tilde g_{5}^2}$.
The consistent set of boundary conditions \cite{CGMPT,CGPT,CCGT} is:
\begin{eqnarray}
&
{\rm at }\  z=R^\prime:
&
\left\{
\begin{array}{l}
\partial_z (g_{5R} A^{L\,a}_\mu + g_{5L} A^{R\,a}_\mu) - \tau M^2
          \frac{R'}{R} (g_{5R} A^{L\,a}_\mu
+ g_{5L} A^{R\,a}_\mu) = 0\,, \\
\tv{15}
g_{5L} A^{L\,a}_\mu  - g_{5R} A^{R\,a}_\mu =0, \
\partial_z B_\mu - \tau' M^2 \frac{R'}{R} B_\mu  = 0\,;
\end{array}
\right.
\label{bc1TeV}\\
&{\rm at }\ z=R:
&
\left\{
\begin{array}{l}
\partial_z A^{L\,a}_\mu + r M^2 A^{L\,a}_\mu=0, \ A^{R\,1,2}_\mu=0\,, \\
\tv{15}
\partial_z (g_{5R}  B_\mu + \tilde g_5  A^{R\,3}_\mu ) + r' M^2 (g_{5R}
         B_\mu + \tilde g_5  A^{R\,3}_\mu )= 0\,, \\
\tv{15}
\tilde g_5 B_\mu - g_{5R} A^{R\,3}_\mu=0\,.
\end{array}
\right.
\label{bc2TeV}
\end{eqnarray}

Thus the parameters of the gauge sector of the theory are given by
$R$, $R'$, $g_{5L}$, $g_{5R}$, $\tilde{g}_{5}$, $r$, $r'$, $\tau$,
$\tau'$.
TeV scale observables are quite insensitive to the precise magnitude of
$R$, as long as it is much smaller than $R'$. One combination of the
remaining parameters is fixed by the $W$ mass, while the matching of the
4D couplings $g$, $g'$ determines two more parameters. Therefore one can
pick as free parameters of the theory the following set: $R$,
$g_{5R}/g_{5L}$, $r$, $r'$, $\tau$, $\tau'$.

\section{Oblique Corrections}
\label{sec:oblique}
\setcounter{equation}{0}

A major stumbling block for non-SUSY alternatives to the SM are the
effects of oblique corrections \cite {ST,CET,biglittle}.
In the following we will use $S$, $T$ and $U$ to fit the
$Z$-pole observables, mainly measured at LEP1.
These three parameters are sufficient for predicting all of those
observables.
In \cite{BPRS}, Barbieri {\it et al.} proposed a new enlarged set of
parameters to also take into account the differential
cross section measurements at LEP2. The only additional information
contained in these parameters is the bound on the
coefficients of the four-Fermi operators that are generated by the
exchange of gauge boson KK modes.
In our language $S$, $T$ and $U$ are a linear combination\footnote{Note
however the slight difference in the definition of the SM couplings
$g$, $g'$: in our approach they are directly defined at $M_Z$, namely
they are the tree level couplings of the mass eigenstate. On the other
hand, in \cite{BPRS} they are defined at low energy, thus also
including contributions of four--fermi operators.} of the
parameters of~\cite{BPRS}.
In our approach we simply use the bounds on $S,T$ and $U$ from the
$Z$-pole observables, while the bounds on the four Fermi
operators are taken into account by directly imposing the constraints
on new gauge bosons from LEP2 and from the direct searches at Tevatron.

Perturbatively the $S$ parameter ``counts" the number of degrees of
freedom that participate in
the electroweak sector, while the $T$ parameter measures the amount of
additional isospin breaking. Contributions to $U$ are typically very
small. Both $S$ and $T$
must be typically small
($< 0.25$) in order
to be compatible with precision electroweak measurements \cite{STUexp}.
To be more precise, in a Higgsless model we should
compare with a fit that assumes a large Higgs mass, namely equal to the
cutoff of the theory\footnote{It is easy to understand it if we think
of Higgsless models as theories with Higgses that are removed by
sending their VEVs (and masses) to infinity.}. In this case, a slightly
negative
$S$ and positive $T$ are preferred~\cite{BPRS}.

\subsection{Planck brane localized fermions.}
%

\begin{figure}[tb]
\begin{center}
\includegraphics[width=11cm]{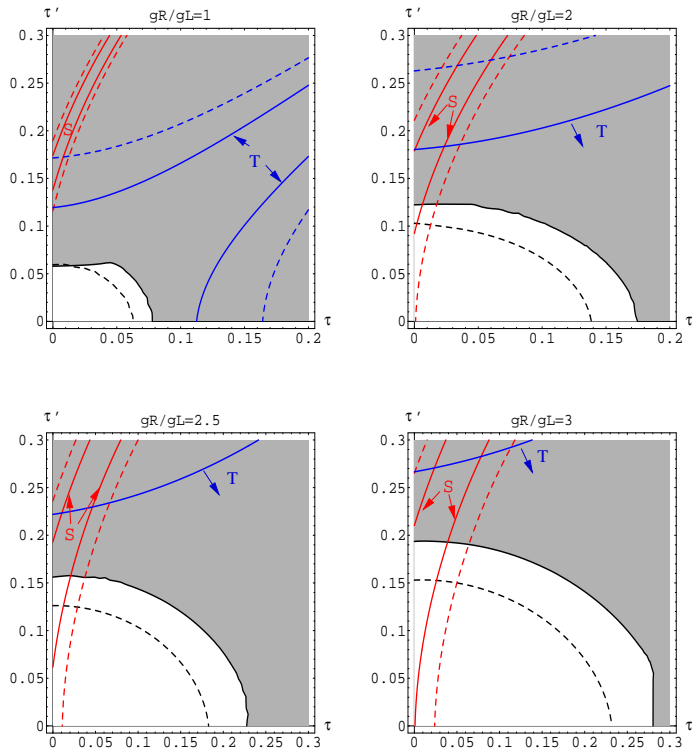}
\end{center}
\caption{Combined plots of the experimental constraints on Higgsless
models for different values of the $g_{5R}/g_{5L}$ ratio, in the
parameter space $\tau$--$\tau'$ (normalized by $R \log R'/R$).
The solid contours for $S$ (red) and $T$ (blue) are at $0.25$; the
dashed contours at $0.5$.
The black solid (dashed) line corresponds to a deviation in the
differential cross section of $3\%$ ($2\%$) at LEP2.
The shaded region is excluded by a deviation larger that $3\%$ at LEP
and/or direct search at Run1 at Tevatron.} \label{fig:combo}
\end{figure}

Electroweak symmetry breaking sectors that are more complicated than
a 4D Higgs doublet tend to have positive $S$ parameters of order 1.
In Higgsless models with
a warped extra dimension it has been shown \cite{CCGT} that both the
ratio of $SU(2)_L$ and $SU(2)_R$ couplings, $g_{5R}/g_{5L}$ as well as
kinetic terms on the TeV brane affect the $S$ and $T$ parameters
in important ways. With no brane kinetic terms and $g_{5R}=g_{5L}$,
$S=1.15$
and $T=0$. Increasing the ratio $g_{5R}/g_{5L}$ reduces\footnote{This is
in accord with the crude expectations for a chiral technicolor theory
\cite{chiraltc}.} $S$ to
\beq
S \approx \frac{6 \pi}{g^2 \log \frac{R^\prime}{R}}
\frac{2}{1+\frac{g_{5R}^2}{g_{5L}^2}}
\eeq
while keeping $T\approx 0$.
A qualitatively similar effect is induced by Planck brane kinetic
terms, the only difference being in the couplings of the gauge bosons,
thus affecting the bounds on direct $Z'$ searches.
As shown previously in~\cite{CCGT} the TeV brane kinetic terms
produce further corrections. The non-Abelian
brane kinetic term gives a correction to $S$ at first order,
multiplying the previous result by $1+ \sfrac{4}{3} \frac{\tau}{R}$,
while giving
a very small positive contribution to $T$.
The $\tau^\prime$ corrections are more complicated, and more
interesting.  The first effects appear
at quadratic order, and they give negative corrections to both $S$ and
$T$. The Abelian
brane kinetic term, $\tau^\prime$, also has the effect of reducing the
mass of the lightest neutral
KK gauge boson resonance.
We scanned the model in this 3D parameter space,
$(g_{5R}/g_{5L},\tau,\tau^\prime)$, to uncover regions allowed by
experiments.
In Fig.\ref{fig:combo} we show combined plots for four values of
$g_{5R}/g_{5L}$= 1, 2, 2.5, 3.
In order to satisfy both precision tests and LEP2/Tevatron bounds, a
large $g_{5R}/g_{5L}$ ratio is required.
In this case, however, the masses of the resonances are raised,
making them possibly ineffective in restoring partial wave unitarity
and leading to strong coupling below $2$~TeV.
These results are in agreement with the conclusions of
refs.~\cite{BPRS,DHLR2}.

\subsection{Delocalized Reference Fermions}
\label{sec:deloc}

In the following, we would like to focus on an
alternative solution to the $S$  problem which has additional
beneficial side-effects.
It has been known for a long time in Randall-Sundrum (RS) models with a
Higgs that the
effective $S$ parameter
is large and negative \cite{CET}  if the fermions are localized on the
TeV brane as
originally proposed \cite{RS}. When the fermions
are localized on the Planck brane the contribution to $S$ is positive,
and so for some intermediate localization the $S$ parameter vanishes,
as first pointed out for RS models by Agashe et al.\cite{ADMS}. The
reason for this is fairly simple.  Since the $W$ and $Z$ wavefunctions
are approximately flat, and the gauge KK mode wavefunctions are
orthogonal to them, when the fermion wavefunctions are
also approximately flat the overlap of a gauge KK mode with two
fermions will approximately vanish. Since it is the coupling of the
gauge KK modes to the fermions that induces a shift in the $S$
parameter, for approximately flat fermion wavefunctions the $S$
parameter must be small.
Note that not only does reducing the coupling to gauge KK modes reduce
the $S$ parameter, it also weakens the experimental constraints on the
existence of light KK modes.
This case of delocalized bulk fermions is not covered by the no--go
theorem of~\cite{BPRS}, since there it was assumed that the fermions
are localized on the Planck brane.

In order to quantify these statements, it is sufficient to
consider a toy model where all the three families of fermions are
massless and have a universal delocalized profile in the bulk.
We first briefly review the bulk equation of motion in AdS$_5$. In 5D
fermions are vector-like, so that they contain both a left- and
right-handed component:
\beq
\Psi = \left( \begin{array}{c}
\chi \\
{\psi}
\end{array} \right)~,
\eeq
where the boundary conditions can be chosen such that there is a zero
mode either in the left--handed (lh) or in the right--handed (rh)
component.
Taking into account the AdS$_5$ metric and spin
connection~\cite{CGHST}, the bulk Lagrangian is the following:
\begin{equation}
S = \int d^5 x
\left(\frac{R}{z}\right)^4
     \left(
- i \bar{\chi}  \bar{\sigma}^\mu \partial_\mu \chi
- i \psi  \sigma^{\mu} \partial_\mu \bar{\psi}
+ \sfrac{1}{2} ( \psi \overleftrightarrow{\partial_z} \chi
-  \bar{\chi}  \overleftrightarrow{\partial_z} \bar{\psi} )
+ \frac{c}{z} \left( \psi \chi + \bar{\chi} \bar{\psi} \right)
\right),
\end{equation}
where $c$ is a bulk Dirac mass in unit of the AdS curvature $1/R$.
The bulk equations of motion derived from this action are:
\begin{eqnarray}
\label{bulkeq1}
-i \bar{\sigma}^{\mu} \partial_\mu \chi - \partial_z \bar{\psi} +
\frac{c+2}{z} \bar{\psi} = 0,
     \\
-i \sigma^{\mu} \partial_\mu \bar{\psi} + \partial_z \chi +
\frac{c-2}{z} \chi = 0.
\label{bulkeq2}
     \end{eqnarray}

If the zero mode is lh, the solution is the following:
\begin{equation}
\chi_0 = A_0  \left( \frac{z}{R} \right)^{2-c_L},
\end{equation}
where the normalization is fixed by the condition
\begin{equation}
            \label{eq:normM0}
\int_R^{R'} dz
\left(\frac{R}{z} \right)^5 \frac{z}{R} \, A_0^2  \left(  \frac{z}{R}
\right)^{4-2c_L}
=1\,,
\ \ i.e. \ \
A_0= \frac{\sqrt{1-2c_L}}{R^{c_L}\sqrt{R^{\prime\,
1-2c_L}-R^{1-2c_L}}}~.
\end{equation}

A similar result applies for rh solutions, where $c_L$ is replaced by
$-c_R$.
Studying the above profile, it's easy to show that lh (rh) fermions are
localized on the Planck brane if $c_L>1/2$ ($c_R<-1/2$), else on the
TeV brane, while for $c_L=1/2$ ($c_R=-1/2$) the profile is flat.

Now, the gauge couplings of the fermions will depend on the parameter
$c$ through the bulk integral of the gauge boson wave functions.
For a lh fermion, that transforms under the bulk gauge group as a $2_L
\times 1_R \times q_{B-L}$ representation, it reads:
\begin{equation}
a_0\, Q\, \gamma_\mu + g_{5L}\, \mathcal{I}_l^{L\mp} (c_L)\, T_{L\pm}
W^{\mp}_\mu + g_{5L}\,
\mathcal{I}_l^{(L3)} (c_L)\, \left( T_{L3} + \frac{\tilde g_5\,
\mathcal{I}_l^{(B)} (c_L)}{g_{5L}\,
\mathcal{I}_l^{(L3)} (c_L)}\, \frac{Y}{2} \right) Z_\mu~,
\label{eq:couplings}
\end{equation}
where we have used that $Y/2 = Q_{B-L}$ (for $SU(2)_R$ singlets) and
the electric charge is defined as $Q=Y/2 + T_{L3}$, and:
\begin{equation}
\mathcal{I}^X_l (c) = A_0^2 \, \int_R^{R'} dz
\left(\frac{R}{z} \right)^{2\, c} \,  \phi_1^X (z)~.
\end{equation}
Here, following the notation in \cite{CCGT}, $a_0$ is the photon wave
function, while $\phi_1^X (z)$ are the $W$ or $Z$ profiles.

Eq.~(\ref{eq:couplings}) is a generalization of eq.~(3.1) in
\cite{CCGT},
where the value of the gauge boson wave functions on the Planck brane
is replaced by the bulk integral, weighted by the fermion profile
squared.
Only the electric charge does not depend on the fermion profile, as the
massless photon is flat along the extra dimension.
However, such corrections to the gauge couplings are universal, so they
can be cast into the definition of the oblique parameters and yield an
effective shift of $S$, $T$ and $U$.

In order to do that, we have to impose the following matching condition
between the 4D couplings and the 5D parameters of the
theory\footnote{Note that this equation does not depend on the overall
normalization of the $Z$ wave function, but is completely determined by
the boundary conditions in eqs.~(\ref{bc1TeV}--\ref{bc2TeV}).}:
\beq \label{tgdef}
\tan^2 \theta_W = \frac{{g'}^2}{g^2} = - \frac{\tilde g_5\,
\mathcal{I}_l^{(B)} (c_L)}{g_{5L}\,
\mathcal{I}_l^{(L3)} (c_L)}~,
\eeq
while the matching of the electric charge remains unaffected:
\beq \label{edef}
\frac{1}{e^2} = \frac{1}{a_0^2}=  \left(
\frac{1}{\tilde{g}_5^2}+\frac{1}{g_{5L}^2}+\frac{1}{g_{5R}^2}
\right) R \log \frac{R'}{R}~.
\eeq

Analogously, the $W$ and $Z$ wave function normalizations are
determined by the following equations:
\beq
g_{5L}\, \mathcal{I}_l^{(L\pm)} (c_L) & = & g~,\\
g_{5L}\, \mathcal{I}_l^{(L3)} (c_L) &=& g \cos \theta_W~.
\eeq
All the oblique corrections are now contained in the wavefunction and
mass renormalizations of the gauge bosons.

For a rh fermion, that transforms as $1_L \times 2_R \times q_{B-L}$,
the situation is a little bit more complicated.
The gauge couplings are the following:
\begin{equation}
a_0\, Q\, \gamma_\mu +  \tilde g_5\, \mathcal{I}_r^{B} (c_R)\,
\frac{Y}{2} Z_\mu  + g_{5R}\, \mathcal{I}_r^{R\mp} (c_R)\, T_{R\pm}
W^{\mp}_\mu + \left( g_{5R}\,
\mathcal{I}_r^{(R3)} (c_R) - \tilde g_5\, \mathcal{I}_r^{(B)} (c_R)
\right)\, T_{R3} Z_\mu~,
\label{eq:couplingsR}
\end{equation}
where
\begin{equation}
\mathcal{I}^X_r (c) = \mathcal{I}^X_l (-c)~.
\end{equation}
In the simple case where $c_R=-c_L$, the first two terms of
eq.~(\ref{eq:couplingsR}) match the SM gauge couplings as defined above.
However, the two additional terms vanish on the Planck brane only, due
to the boundary conditions.
In general they will give rise to non-oblique corrections (see
sec.~\ref{sec:light}).

The computation of the oblique corrections follows straightforwardly
from the matching conditions (\ref{tgdef}) and (\ref{edef}).
Before showing some numerical results, it is useful to understand the
analytical behavior of $S$ in interesting limits.
For fermions almost localized on the Planck brane, it is possible to
expand the result in powers of $(R/R')^{2c_L-1} \ll 1$.
The leading terms, also expanding in powers of $1/\log$, are:

\beq
S = \frac{6 \pi}{g^2\, \log \frac{R'}{R}} \left( 1 - \frac{4}{3}
\frac{2c_L-1}{3-2c_L} \left( \frac{R}{R'}\right)^{2c_L-1} \log
\frac{R'}{R} \right)~,
\eeq \label{eq:Splanck}
and $U \approx T \approx 0$.
The above formula is actually valid for $1/2 < c_L < 3/2$. For
$c_L>3/2$ the corrections are of order $(R'/R)^2$ and numerically
negligible.
As we can see, as soon as the fermion wave function starts leaking into
the bulk, $S$ decreases.

Another interesting limit is when the profile is almost flat, $c_L
\approx 1/2$.
In this case, the leading contributions to $S$ are:

\beq
S =  \frac{2 \pi}{g^2\, \log \frac{R'}{R}} \left( 1 + (2 c_L -1)\, \log
\frac{R'}{R} + \mathcal{O} \left((2c_L-1)^2 \right)\right)~.
\eeq
In the flat limit $c_L=1/2$, $S$ is already suppressed by a factor of 3
with respect to the Planck brane localization case.
Moreover, the leading terms cancel out for:
\beq
c_L = \frac{1}{2} - \frac{1}{2\, \log \frac{R'}{R}} \approx 0.487~.
\eeq

For $c_L<1/2$, $S$ becomes large and negative and, in the limit of TeV
brane localized fermions ($c_L \ll 1/2$):

\beq
S =  - \frac{16 \pi}{g^2} \frac{1-2 c_L}{5-2 c_L}~,
\eeq
while, in the limit $c_L\rightarrow - \infty$:
\begin{eqnarray}
T&\rightarrow& \frac{2 \pi}{g^2\, \log \frac{R'}{R}} (1 + \tan^2
\theta_W) \approx 0.5~,\\
U&\rightarrow& - \frac{8 \pi}{g^2\, \log \frac{R'}{R}} \frac{\tan^2
\theta_W}{2 + \tan^2 \theta_W} \frac{1}{c_L} \approx 0~.
\end{eqnarray}

\begin{figure}[tb]
\begin{center}
\includegraphics[width=12cm]{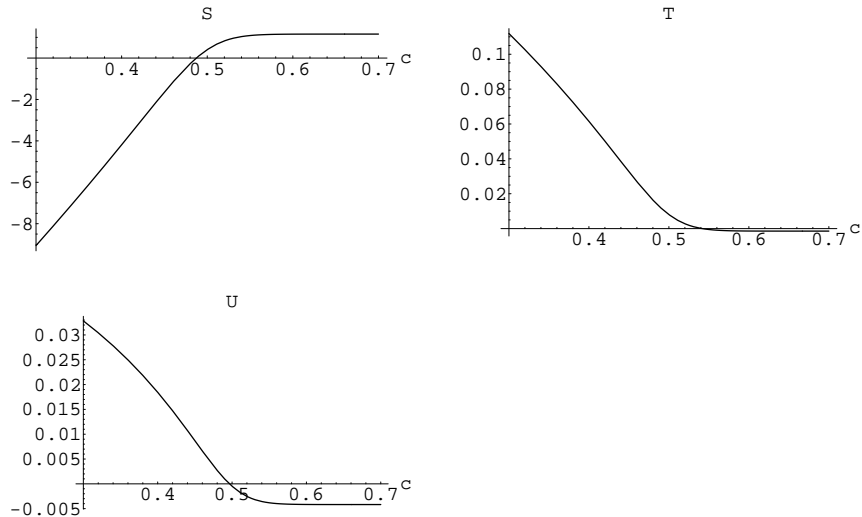}
\end{center}
\caption{
Plots of the oblique parameters as function of the bulk mass of the
reference fermion. The values on the right correspond to localization
on the Planck brane. $S$ vanishes for $c=0.487$.
} \label{fig:STUvsC}
\end{figure}

In Fig.~\ref{fig:STUvsC} we show the numerical results for the oblique
parameters as function of $c_L$.
We can see that, after vanishing for $c_L \approx 1/2$, $S$ becomes
negative and large, while $T$ and $U$ remain smaller.

\section{Perturbative Unitarity} \label{sec:unitarity}
\setcounter{equation}{0}
\setcounter{footnote}{0}

The other criticism leveled at Higgsless models is that if the KK modes
are above $1$~TeV then perturbative unitarity might break down  near
$2$~TeV,
thus rendering perturbative calculations impossible.  If this were
the case then Higgsless theories would look very much like technicolor
theories and it would be difficult to make any theoretical progress.

The amplitude of elastic scattering of longitudinal massive $W$ blows
up at around $1.8$~TeV, thus violating perturbative unitarity.
In the SM it is restored by the contribution of the Higgs field, that
cancels the residual term growing with the energy squared.
As is by now well known, in Higgsless models perturbative unitarity
breakdown is delayed by the contribution of the KK modes of ordinary
gauge bosons~\cite{CGMPT}: as a consequence of 5D gauge invariance, two
sum rules involving trilinear gauge couplings and masses ensure the
cancellation of the terms growing like $E^4$ and $E^2$.
This implies a first constraint on the spectrum.
Over large regions of the parameter space the sum
rules are very accurately satisfied with only the first two KK modes,
which typically happens when masses stay below $1500$~GeV.

Even though the growing terms are tamed, it is still possible for the
tree--level elastic scattering amplitude to break down around $2-3$~TeV
in
Higgsless theories~\cite{DHLR1,DHLR2}.
However, this residual growth is due to the presence of large
logarithms coming from the forward scattering region, after integrating
over the
scattering angle.
Such logarithms are present in the SM as well (in particular from the
t-channel photon exchange, whose contribution is divergent), but in our
case they are enhanced by a large coefficient growing with the
resonance mass.
The log term in the s--partial wave amplitude is given by:
\beq
\frac{g_{WW Z_{k}}^2}{32 \pi}
\left( 2 - \frac{M_{Z_{k}}^2}{M_W^2}\right)^2
\log\left(\frac{4 E^2}{M_{Z_{k}}^2}\right)~.
\label{largelog}
\eeq
Such behavior is not particular to Higgsless theories and can be easily
reproduced in simple 4D toy models.
In this case however they are unphysical: they arise when the
energy is much larger than the resonance mass, where other inelastic
channels open up and cannot be neglected in the unitarization of the
$S$ matrix~\cite{Papucci}.
Thus, the spoiling of partial wave unitarity, if due to such large
logs, cannot indicate the scale where a strong coupling regime is
entered.

On the other hand, from a 5D point of view a linear growth of the
amplitude is  expected.
Indeed, according to naive dimensional analysis (NDA), the loop factor
grows with the energy as
\beq
\frac{g_5^2}{24 \pi^3} E\,.
\eeq
    From the strength of this loop factor, warped down to the TeV scale,
we
conclude that perturbativity breaks down  around a scale
\beq \label{eq:NDA}
\Lambda_{\rm NDA} \sim \frac{24 \pi^3}{g_5^2} \frac{R}{R'}\,.
\eeq
In the warped Higgsless model, the NDA cutoff scale can be expressed in
terms of the masses of the $W$ and the first KK excitation and the 4D
SM gauge coupling:
\beq
\Lambda_{\rm NDA} \sim \frac{12 \pi^4 M_W^2}{g^2 M_{W^{(1)}}}\,.
\eeq
    From the formula above, it is clear that the heavier the resonance,
the
lower the scale where perturbative unitarity gets lost.
This also gives a rough estimate, valid up to a numerical coefficient,
of the actual scale of non--perturbative physics.
An explicit calculation of the scattering amplitude, including
inelastic channels, shows that this is indeed the case and the
numerical factor is found to be roughly $1/4$~\cite{Papucci}.

Since the ratio of the $W$ to the first KK mode mass squared
is of order
\beq
\frac{M_W^2 }{M_{W^{(1)}}^2}= \mathcal{O}
\left( {1}/{\log  \left({R^\prime}/{R}\right)} \right)~,
\eeq
raising the value of $R$ (corresponding to lowering the 5D UV scale)
will significantly increase the NDA cutoff.
With $R$ chosen to be the inverse Planck scale, the first KK resonance
appears around $1.2$~TeV, but for larger values of $R$ this scale can
be safely reduced down below a TeV.
As already discussed in the previous section, such resonances will be
weakly coupled to almost flat fermions and can easily avoid the strong
bounds from direct searches at LEP or Tevatron.
If we are imagining that the AdS space is a dual description
of an approximate conformal field theory (CFT), then $1/R$ is the scale
where the CFT is no longer
approximately conformal and perhaps becomes asymptotically free. Thus it
is quite reasonable that the scale $1/R$ would be much smaller than the
Planck scale.

\begin{figure}[tb]
\begin{center}
\includegraphics[width=10cm]{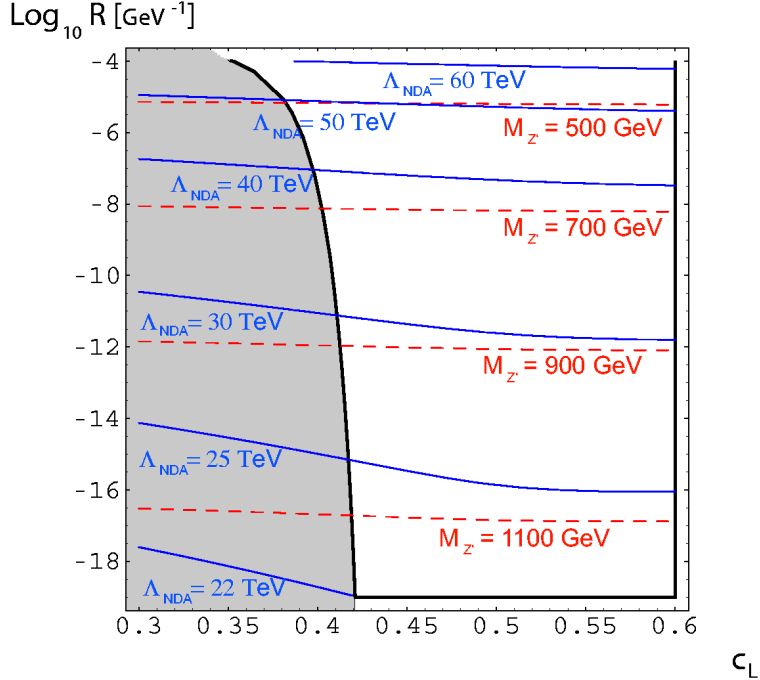}
\end{center}
\caption{Contour plots of $\Lambda_{\rm NDA}$ (solid blue lines) and
$M_{Z^{(1)}}$ (dashed red lines) in the parameter space $c_L$--$R$.
The shaded region is excluded by direct searches of light $Z^\prime$ at
LEP.
}
\label{fig:NDA}
\end{figure}

\begin{figure}[tb]
\begin{center}
\includegraphics[width=14cm]{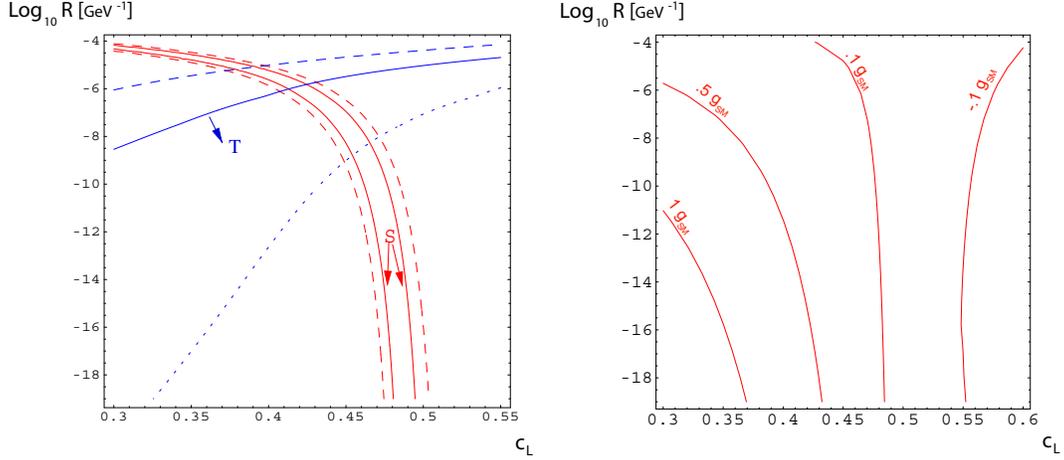}
\end{center}
\caption{
On the left, contours of $S$ (red), for $|S|=0.25$ (solid) and $0.5$
(dashed) and $T$ (blue), for  $|T|=0.1$ (dotted), $0.3$ (solid) and
$0.5$ (dashed), as function of $c_L$ and $R$.
On the right, contours for the generic suppression of fermion couplings
to the first resonance with respect to the SM value. In particular we
plotted the couplings of a lh down--type massless quark with the $Z'$.
The region for $c_L$, allowed by $S$, is between $0.43\div 0.5$, where
the couplings are suppressed at least by a factor of 10.}
\label{fig:coupR}
\end{figure}

In Fig.~\ref{fig:NDA} we have plotted the value of the NDA scale
(\ref{eq:NDA}) as well as the mass of the first resonance in the
$(c_L-R)$ plane.
Increasing $R$ also affects the oblique corrections.
However, while it is always possible to reduce $S$ by delocalizing the
fermions, $T$ increases and puts a limit on how far $R$ can be raised.
One can also see form Fig.~\ref{fig:coupR} that in the region where
$|S|<0.25$, the coupling of the first resonance with the light fermions
is generically suppressed to less than $10\%$ of the SM value.
This means that the LEP bound of $2$~TeV for SM--like $Z^\prime$ is
also decreased by a factor of 10 at least (the correction to the
differential cross section is roughly proportional to $g^2/M_{Z'}^2$).
In the end, values of $R$ as large as $10^{-7}$~GeV$^{-1}$ are allowed,
where the resonance masses are around $600$~GeV.
So, even if, following the analysis of~\cite{Papucci}, we take into
account a factor of roughly $1/4$ in the NDA scale, we see that the
appearance of strong coupling regime can be delayed up to $10$~TeV.  At
the LHC it will be very difficult to probe $WW$ scattering above 3 TeV.

\section{Flavor Physics}
\setcounter{equation}{0}
\setcounter{footnote}{0}

As already mentioned, fermions masses can be easily reproduced by
boundary conditions~\cite{CGHST}.
Since they must be bulk fields in a Higgsless model (otherwise it
would not be possible to produce a isospin breaking mass spectrum for
them), and since
in 5D the smallest representation of the Lorentz group is a Dirac
spinor,
one is forced to introduce a full Dirac spinor for every SM field.
For the third generation quarks for example this implies that one has
at least
the fields
\begin{equation}
\left( \begin{array}{c} \chi_{tL}\\ {\psi}_{tL} \\  \chi_{bL} \\
{\psi}_{bL}
\end{array} \right) \ \ \
\left( \begin{array}{c} \chi_{tR} \\ {\psi}_{tR} \\  \chi_{bR} \\
{\psi}_{bR}
\end{array} \right),
\end{equation}
where the $\psi$'s are right handed 4D Weyl spinors, while the $\chi$'s
are
left handed 4D Weyl spinors, and the subscript $L,R$ denote whether
these
fields are part of the $SU(2)_L$ or $SU(2)_R$ doublet. In order to get
the correct spectrum, one needs to make sure that the boundary
conditions
of the $L$ and $R$ fields are different, for example by imposing $(+,+)$
boundary conditions on the $\chi_{tL,bL}$ and $\psi_{tR,bR}$ fields, in
order to obtain approximate zero modes,
and consequently applying the opposite $(-,-)$ boundary conditions to
the remaining fields.

An acceptable mass spectrum can then be generated by noting that the
gauge
group on the TeV brane is non-chiral, and therefore a Dirac mass
\begin{equation}
M_D R' (\chi_{tL} \psi_{tR} +\chi_{bL} \psi_{bR})
\end{equation}
can be added on the TeV brane. Due to the remaining $SU(2)_D$ gauge
symmetry
the same term has to be added for top and bottom quarks. The necessary
splitting between top and bottom can then be achieved by adding a large
brane induced kinetic term for $\psi_{bR}$ on the Planck brane (where
$SU(2)_R$ is broken)~\cite{BPR}. This is equivalent to adding a mixing
on the
Planck
brane to a localized singlet field, that we parametrize by the ratio
$\xi$ between the mixing mass and the mass of the localized
field~\cite{CGPT,CGHST}.

In the following subsections we will address some issues about flavor
physics arising in this scenario.
First of all, the eventual presence of flavor changing neutral currents
(FCNC) induced either by higher dimension operators or by
non--universal corrections.
Next, we will briefly discuss the problems surrounding the inclusion of
the third family of quarks in the picture. For interesting flavor  
physics
signals in warped extra dimensional models see~\cite{APS}.

\subsection{FCNC from higher order operators} \label{sec:FCNC}

In a warped background, the scale suppressing dimensionful operators
depends on the position of the fields along the
extra-dimension~\cite{GP}: for operators involving fields mostly
localized on the UV brane, the suppression scale will be approximately
$1/R$, while for operators with fields localized on the IR brane, the
scale will rather be around $1/R'$. There are severe constraints on the
scale of four-Fermi operators leading to FCNC, putting a lower bound
around~$10^3$~TeV. While this constraint was clearly satisfied when the
two light generations of quarks and leptons were localized close to the
Planck brane,
it becomes more worrisome when the fermions are delocalized in the
bulk, a situation favored as we have just seen by electroweak precision
measurements.

Let us for instance consider the 5D operator
\beq
	\label{eq:5D-FCNC}
\int d^5x \left(\frac{R}{z}\right)^5 \, R^3\ \bar{\Psi}_L \Gamma^M
\Psi_L \ \bar{\Psi}_L \Gamma_M \Psi_L
\eeq
where $\Gamma^M$ are the 5D Dirac matrices (see for instance app. A of
\cite{CGHST}) and where $\Psi_L$ is the 5D SU(2)$_L$ doublet of the
first (or second) generation of quarks and containing  in particular
the $u_L$ and $d_L$ zero modes. Note that the scale suppressing this 5D
operator is set by the 5D UV cutoff $1/R$. Upon compactification, this
operator will in particular generate the 4D FCNC operator
\beq
\int d^4x\ \frac{1}{\Lambda_{FCNC}^2} \ \bar{\chi}_{u_L} \gamma^\mu
\chi_{u_L}
\ \bar{\chi}_{d_L} \gamma_\mu \chi_{d_L}
\eeq
where the scale $\Lambda_{FCNC}$ is obtained from eq.~(\ref{eq:5D-FCNC})
after integration of the fermion zero-mode profiles over the
extra-dimension. For $c_L  \approx 1/2$, we get:
\beq
\Lambda_{FCNC}^2 \approx \frac{(R/R')^{2-4c_L} \log (R'/R)}{R'^2}.
\eeq
For $1/R'\sim 1$~TeV, to get a suppression factor of $10^3$~TeV, $c_L$
would have to be bigger than $0.57$.
Clearly the values of $c_L$ used to reduce the $S$ parameter do not
fulfill this criterion, which means that the set-up fails to naturally
explain the absence of FCNC and additional flavor symmetries in 5D
would be necessary. It is however relatively easy to impose such a
flavor symmetry in the bulk and on the TeV brane
and naturally break it close to the Planck brane. Due
to the small overlap of the fermion wavefunctions on the Planck brane,
the suppression scale of the four-Fermi operators will be significantly
increased.

Finally let us mention that other 5D operators like
\beq
\int d^5x \left(\frac{R}{z}\right)^5 \, R^3\ \bar{\Psi}_L \Psi_R \
\bar{\Psi}_R \Psi_L
\eeq
that would lead to the 4D operator
\beq
\int d^4x\ \frac{1}{\Lambda_{FCNC}^2} \ \chi_{u_L}  \psi_{d_R}
\ \psi_{d_R}  \chi_{u_L}
\eeq
are less constraining since the suppression scale $\Lambda_{FCNC}$ can
be now raised by localizing the rh components of the fermion on Planck
brane ($c_R<-1/2$), as discussed in sec.~\ref{sec:light}. Thus  for
these operators $\Lambda_{FCNC}$ can be as high as $10^3$~TeV even if
$c_L<1/2$.

\subsection{Non--oblique Corrections: Light Fermions.} \label{sec:light}

Due to fermions propagating into the bulk, the model is also affected
by corrections to the gauge couplings that cannot be removed by  a
shift of the oblique parameters.
There are two types of such corrections:  corrections coming from the
enlarged gauge structure in the bulk (already mentioned in
sec.~\ref{sec:deloc}), and non--universal corrections coming from the
different fermion masses. 	
We can parametrize the combination of these corrections as shifts with
respect to the SM couplings, as follows:
\beq
\begin{array}{lcl}
g^Z_{f_l} = \left( 1+\gamma_l^f \right) \frac{g}{\cos \theta_W} \left(
T_3 + \sin^2 \theta_W\, Q \right)\,, & \quad & g^W_{f_l} = \left(
1+\omega_l^f \right) g\,,\\
g^Z_{f_r} =\left( 1+\gamma_r^f \right) g' \sin \theta_W\, Q\,, & \quad
& g^W_{f_r} =\omega_r^f  g\,.
\end{array}
\eeq

     The  corrections arising from the enlarged gauge structure affect  
the
couplings of the rh fermions\footnote{Note that this particular
structure comes from our choice for the matching condition of the 4D
gauge couplings. For example, another possible choice would be to match
the SM couplings with the couplings of the up--type fermions. In this
case, there would be non--oblique corrections  affecting the couplings
of down--type fermions.}. They are present  in the massless limit, and
are universal provided that  the bulk masses of the $SU(2)_R$ doublets,
$c_R$, are all equal.
These corrections are generically not very tightly bounded by
experiment.
For example, from the  $\mu$ decay the limits on the $W$ couplings with
rh electron and muon, $\omega_r^{e,\,\mu}$, is of order few \% with
respect to the SM gauge coupling, while from the $\tau$ leptonic decays
$\omega_r^\tau$ has to be smaller than 10\%~\cite{PDG}.

On the other hand, boundary conditions generating fermion masses
distort the zero mode profiles, introducing non-universal corrections
to the gauge couplings.
The more the fermion probes the bulk, the more severe such
non--universalities are, as the wave function will be more sensitive to
the
TeV-localized mass term.
For the first generation, the corrections given by the masses (much
smaller than the electroweak scale) are negligible.
Nevertheless, there are non--oblique corrections that generically will
be $\gamma_l \approx \omega_r \approx \mbox{few}\cdot 10^{-4}$ for
$c_R<-1/2$.

The most stringent experimental constraints come from
non--universalities between the first two generations of quarks, for
example $\delta \gamma^{ds}_l = \gamma_l^s - \gamma_l^d$ and $\delta
\gamma^{ds}_r = \gamma_r^s - \gamma_r^d$ from Kaon
physics~\cite{Burdman}.
The bounds generically imply $| \delta \gamma^{ds} | < 10^{-5}$.
Weaker bounds also come from $B^0$--$\bar B^0$ and $D^0$--$\bar D^0$
mixings, namely $| \delta \gamma^{cu} | < 10^{-4}$ and $| \delta
\gamma^{bd} | < 5\cdot 10^{-4}$.
Such bounds depend on unknown quark mixing matrices, so can be weaker
if small elements are involved (see~\cite{Burdman} for more details).
It is possible, however, to tune $c_L$ and $c_R$ for the second
generation in order to fulfill such bounds.
A numerical example is shown in table \ref{tab:nonuniv}, in the minimal
scenario where only a localized kinetic term is added to the
R-component of the lighter quark.
In this example $\gamma_r^c$ is too large.
It is however easy to suppress it, for example splitting the $s_r$ and
$c_r$ into two different bulk $R$--doublets with different bulk masses,
$c_{Rs}$ and $c_{Rc}$.
Similar arguments apply to the leptons.

\begin{table}[th]
\begin{center}
\begin{tabular}{||l|c|c|c|c||}
\hline
\hline
u & $\gamma_l^u < 10^{-6}  $ & $\omega_l < 10^{-6}$ & $\gamma_r^u =
-3.95\cdot 10^{-4}$ & $\omega_r = 1.33\cdot 10^{-5} $ \\
d & $\gamma_l^d < 10^{-6} $ &  & $\gamma_r^d = -5.85\cdot 10^{-4} $ & \\
\hline
s & $\gamma_l^s = -8\cdot 10^{-6}$ & $\omega_l= -5\cdot 10^{-5}$ &
$\gamma_r^s =-5.9\cdot 10^{-4} $ & $\omega_r =3.8 \cdot 10^{-4} $ \\
c & $\gamma_l^c = -3.6\cdot 10^{-6} $ &  & $\gamma_r^c = -1.5\cdot
10^{-2}$ & \\
\hline
\hline
\end{tabular}
\caption{Parameters used in this example: $c_{L1}=0.485$,
$c_{R1}=-0.6$, $M_{D1}=3.17$\, GeV, $\xi_{u_r}=4.71$;
$c_{L2}=0.485448$, $c_{R2}=-0.511$, $M_{D2}=47.5$\, GeV, $\xi_{s_r}=
59.5$; where $\xi$ is the parameter describing the localized kinetic
term. In this case $\gamma_r^s - \gamma_r^d = 6\cdot 10^{-6}
$.}\label{tab:nonuniv}
\end{center} \end{table}

As already discussed in sec.~\ref{sec:FCNC}, in the above scenario
generically large FCNC are expected to arise from higher dimensional
operators, as flavor symmetry is broken both by the Dirac masses on the
TeV brane and by the bulk masses.
An elegant way out is to impose a bulk flavor symmetry, and exile all
the flavor structure on the Planck brane.
In this case, higher dimensional operators will be generically safely
suppressed by the scale $1/R$.
In a minimal scenario, one can add a universal Dirac mass for quarks
(and another for leptons) in order to account for the heaviest particle
(charm or $\tau$).
Then the lighter masses are suppressed by localized kinetic terms for
the $L$--doublet and the two singlets on the Planck brane.

\subsection{Top mass, $Zb\bar{b}$ coupling and loop induced isospin
violations}

The major challenge facing Higgsless models is the incorporation of the
third family of quarks.
There is a tension~\cite{BN,ADMS} in obtaining a large top quark mass
without
deviating from the observed bottom couplings with the $Z$. It can be
seen in the following way. The top
quark mass is proportional both to the Dirac mixing $M_D$ on the TeV
brane
and the overall scale of the extra dimension set by $1/R'$. For
$c_L\sim 0.5$ (or larger) it is in fact impossible to obtain a heavy
enough top
quark
mass (at least for $g_{5R}=g_{5L}$). The reason is that for $M_DR' \gg
1$
the light mode mass saturates at
\begin{equation}
m_{top}^2 \sim \frac{2}{R'^2 \log \frac{R'}{R}}\,,
\end{equation}
which gives for this case $m_{top}\leq \sqrt{2} M_W$. Thus one needs to
localize the top and the bottom quarks closer to the TeV brane.
However, even
in this case a sizable Dirac mass term on the TeV brane is needed to
obtain a heavy enough top quark. The consequence of this mass term is
the
boundary condition for the bottom quarks
\begin{equation}
\chi_{bR}= M_D R'\, \chi_{bL}.
\end{equation}
This implies that if $M_D R' \sim 1$ then the left handed bottom quark
has a sizable component also living in an $SU(2)_R$ multiplet, which
however
has a coupling to the $Z$ that is different from the SM value. Thus
there
will be a large deviation in the $Zb_L\bar{b}_L$. Note, that the same
deviation will not appear in the $Zb_R\bar{b}_R$ coupling, since
the extra kinetic term introduced on the Planck brane to split  top and
bottom will imply that the right handed $b$ lives mostly in the induced
fermion on the Planck brane which has the correct coupling to the $Z$.

\begin{figure}[tb]
\begin{center}
\includegraphics[width=8cm]{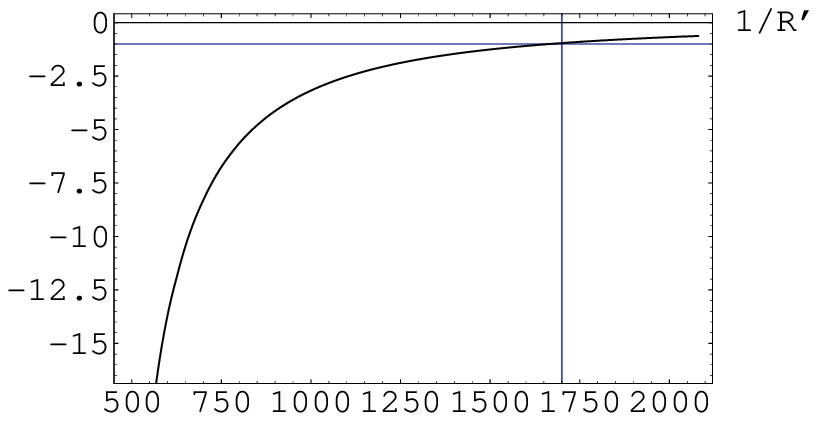} \hspace{0.5cm}
\includegraphics[width=6cm]{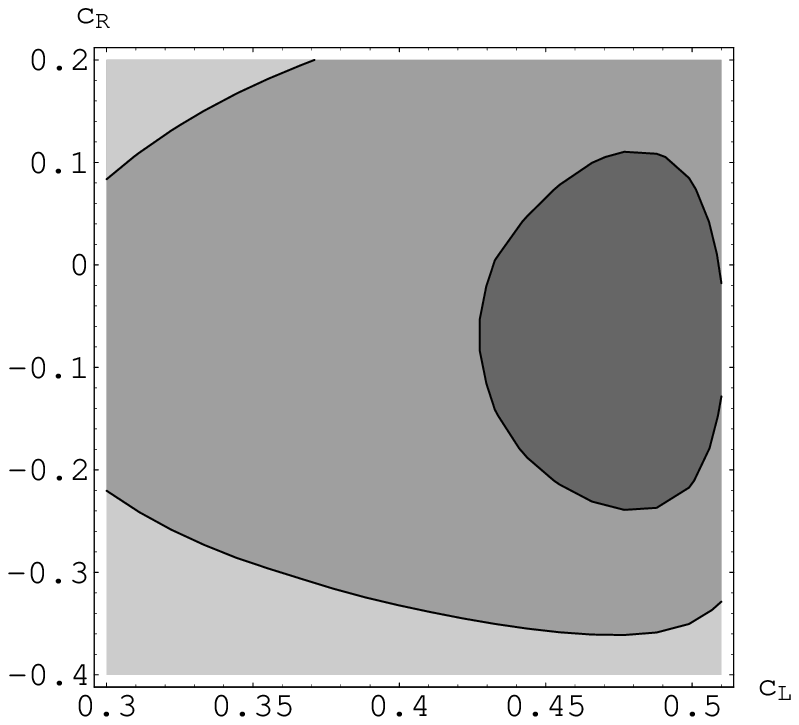}
\end{center}
\caption{
Plots of the percentage deviation of the $Zb_L\bar{b}_L$ coupling with
respect to the SM value as function of the scale $1/R'$ (left, with
$c_L=0.46$ and $c_R=-0.05$) and as function of the bulk masses $c_L$
and $c_R$ for $1/R'=1750$~GeV (right). The contours are at $1\,\%$ and
$1.5\,\%$.
Different values of $1/R'$ are obtained varying the ratio $g_R/g_L$
between 1 and 6; the plot on the right assumes $g_R/g_L=5$.
} \label{fig:zbbbar}
\end{figure}

The only way of
getting around this problem would be to raise the value of $1/R'$, and
thus
lower the necessary mixing on the TeV brane needed to obtain a heavy
top quark. One way of raising the value of $1/R'$ is by increasing
the ratio $g_{5R}/g_{5L}$ (at the price of making also the gauge KK
modes
heavier and thus the theory more strongly coupled). To illustrate
the magnitudes in the deviations of the $Zb_L\bar{b}_L$ coupling we
have plotted the percentage variation with respect to the SM value as a
function of $1/R'$ (varying the ratio).
We can see in the left hand side of Fig.~\ref{fig:zbbbar} that the
deviation decreases with increasing $1/R'$.
In order to be compatible with the experimental bound of
$1\,\%$~\cite{LEP} from LEP, a scale larger that $1700$~GeV is required
(which implies $g_R/g_L > 4.5$ and the first resonance above $4$~TeV),
where the theory is already strongly coupled.
On the right hand side of Fig.~\ref{fig:zbbbar} we also show the
contours of fixed amount of deviation for $g_{5R}/g_{5L}=5$.

Another generic problem arising from the large value of the
top-quark mass in models with warped extra dimensions
comes from the isospin violations in the KK sector of the top and the
bottom quarks. If the spectrum of the
top and bottom KK modes is not sufficiently degenerate, the loop
corrections involving these KK modes
to the $T$-parameter could be large.~\footnote{We thank Kaustubh
Agashe and Roberto Contino for emphasizing the importance of these loop
effects to us.}
  This possibility was pointed out
in~\cite{ADMS}, and further discussed in~\cite{ACP}.
In~\cite{ADMS} an estimate
for the size of these loop corrections was given using a mass insertion
approximation. Since the mass insertions
(the Dirac mixings of the KK modes on the TeV brane) are very large,
possibly larger than the unperturbed masses,
this method likely gives an overestimate of the resulting T-parameter.
In order to get some sense of magnitudes we nevertheless quote
the results found in \cite{ADMS}:
\begin{equation}
T^{top KK} \sim 0.84 \left(\frac{2}{1-2 c_L^{top}}\right)^2 \left(
\frac{m_t}{m_t^{KK}}\right)^2.
\end{equation}
For $c_L$ close to one half and a KK mass for the top of order 700 GeV
this contribution would be enormous.
One can see that in order to suppress this contribution one would again
need to increase $m_t^{KK}$, the first
KK mass of the top quark, which can only be achieved by raising the
value of $1/R'$. One would also need to move the
left handed doublet closer to the TeV brane in order to reduce the
$c_L$ dependent enhancement factor. Both this argument
and the consideration of the $Zb\bar{b}$ vertex would call for a
scenario where the third generation feels a different
value of $R'$ than the rest of the particles. We will speculate about
such a possibility below.

\subsection{Possible future directions for model building}

Given that the main problem with Higgsless models arises from mass
generation of the top and the consequent effects on the couplings of
the $b$ quark, there are several possible directions to explore for
building realistic models.  A simple direction would be to relax the
assumptions that the Higgs VEV is infinite and localized on the TeV
brane.  As long as the Higgs
VEV is large compared to its SM value, its contribution to WW
scattering is suppressed. A  VEV
above 1 TeV is probably sufficient to make its contribution at the LHC
unobservably small.
Once the VEV is finite it is possible to imagine the Higgs having a
profile in the bulk
\cite{LutyTaki}  which will reduce the value of $M_D$ necessary in
order to obtain a large mass, and thus the
mixing between the $L$ and $R$ components of the lh $b$-quarks. In the
dual CFT language this corresponds to the operator which breaks the
electroweak symmetry  having a finite\footnote{Rather than an infinite
dimension corresponding to the strict localization on the TeV brane.}
scaling dimension  $> 2$. This is the type of dynamical breaking
scenario that happens in QCD or technicolor.

      A more ambitious approach would be to separate the fermion and  
gauge
boson Higgs sectors.
      Conceptually we can imagine a theory with two Higgses, one of which
predominantly gives mass to the gauge bosons and one which
predominantly gives mass to the fermions. One could then apply the
Higgs decoupling limit and arrive at a ``double Higgsless" theory. More
concretely we could have two AdS spaces on either side of a Planck
brane, where the gauge bosons can propagate in the entire space and the
fermions can only propagate in the left half. If the ratio of
$g_{5R}/g_{5L}$ gauge couplings and/or the warp factor $R^\prime/R$ are
large on the left-space then electroweak symmetry breaking on the left
      will contribute little to the $W$ and $Z$ gauge boson masses. This
means that the gauge boson
      wavefunctions will be almost flat on the left. Nevertheless, $1/R'$
can be bigger on the left, and in this way the
mixing between the $L$ and $R$ components of the $b$-quark can be
reduced.
Also if $1/R^\prime$ is bigger on the left than the right
      then $KK$ modes of the fermions can be made heavier than the $KK$
modes of the gauge bosons, meaning that problems with Unitarity can be
postponed, while suppressing isospin violating loop corrections from
fermion $KK$ modes \cite{ADMS}.

We plan to study the feasibility of such setups in the future.

\section{Conclusions}
\setcounter{equation}{0}
\setcounter{footnote}{0}

There has recently been a long discussion about the feasibility of
Higgsless models, when facing precision measurements.
The most common criticisms are the large oblique corrections (namely,
a large tree--level contribution to $S$) and a strong coupling
arising from the early breakdown of partial wave unitarity in $W$ boson
scattering.
Due to these problems, some authors have claimed these models to be
disfavored by experiment.
However, we have shown that it is possible to cure these ills by
delocalizing the light fermions in the bulk.
In the limit of almost flat profiles, the gauge boson KK modes almost
decouple from the light fermions, while remaining effective in
restoring perturbative unitarity in $WW$ scattering.
This yields a double advantage: the tree level contribution to $S$ is
suppressed and direct search limits are lowered.
Therefore, a scenario with $600$~GeV resonances and a perturbative
regime up to $10$~TeV is allowed.

Finally, we pointed out that the main challenge still facing Higgsless
models is actually the successful inclusion of a heavy top quark,
without stumbling over large corrections to bottom couplings with the
$Z$.
We have also mentioned some possible future directions in model
building that might lead to a completely realistic model.

\section*{Acknowledgments}
We thank  Neal
Weiner for emphasizing the importance of th $Zb\bar{b}$ coupling. We
thank Kaustubh Agashe
for discussions on the $S$ parameter and the effect of top-bottom KK
mode loops.
We thank Markus Luty and Takemichi Okui for discussion of their
upcoming paper.
We also thank Roberto Contino, Antonio Delgado, Michael Graesser,
Guido Marandella, Michele Papucci, Gilad Perez,
Maxim Perelstein, Alex Pomarol, Riccardo Rattazzi, and James
Wells
for useful discussions and comments.
The research of G.C. and C.C.
is supported in part by the DOE OJI grant DE-FG02-01ER41206 and in part
by the NSF grants PHY-0139738  and PHY-0098631.
C.G. is supported in part by the RTN European Program
HPRN-CT-2000-00148, by the ACI Jeunes Chercheurs 2068
and by the Michigan Center for Theoretical Physics. J.T. is supported
by the US Department of Energy under contract
W-7405-ENG-36.


\end{document}